# Analysis of nulling phase functions suitable to image plane coronagraphy

François Hénault, Alexis Carlotti, Christophe Vérinaud
Institut de Planétologie et d'Astrophysique de Grenoble
Université Grenoble-Alpes, Centre National de la Recherche Scientifique
B.P. 53, 38041 Grenoble – France

**ABSTRACT**

Coronagraphy is a very efficient technique for identifying and characterizing extra-solar planets orbiting in the habitable zone of their parent star, especially when used in a space environment. An important family of coronagraphs is based on phase plates located at an intermediate image plane of the optical system, that spread the starlight outside the "Lyot" exit pupil plane of the instrument. In this communication we present a set of candidate phase functions generating a central null at the Lyot plane, and study how it propagates to the image plane of the coronagraph. These functions include linear azimuthal phase ramps (the well-known optical vortex), azimuthally cosine-modulated phase profiles, and circular phase gratings. Numerical simulations of the expected null depth, inner working angle, sensitivity to pointing errors, effect of central obscuration located at the pupil or image planes, and effective throughput including image mask and Lyot stop transmissions are presented and discussed. The preliminary conclusion is that azimuthal cosine functions appear as an interesting alternative to the classical optical vortex of integer topological charge.

## 1  INTRODUCTION

In the forthcoming years, coronagraphy will probably be the most efficient technique for identifying and characterizing extra-solar planets orbiting in the habitable zone of their parent star, as illustrated by the recent commissioning of the SPHERE [1] and [2] GPI instruments, respectively on the VLT and Gemini South facilities. An extensive review of different types of coronagraphs and of their achievable performance can be found in Refs [3] and [4] respectively. This communication is restricted to the field of Phase-mask coronagraphs (PMC), where phase plates are located at an intermediate image plane of the optical system, and spread the starlight outside the diameter of a "Lyot" exit pupil stop, before final focusing at the image plane of the instrument. In this family of coronagraphs, the circular and four-quadrant PMC probably are the most ancient and well understood [5-6]. However these phase masks are essentially made of a limited number of discontinuous phase steps. Conversely, in the more recent concept of Optical vortex coronagraph (OVC) [7-9] the phase mask exhibits continuous variations along its radial and azimuthal profiles, at the exception of a $2\pi$-phase jump inherent to the creation of the orbital angular momentum [10].

The purpose of this communication is to find other types of phase functions suitable to phase-mask coronagraphy. We only look for analytical functions being continuous, either invariant in azimuth or centro-symmetric. We proceed by similarity with Ref. [11], where a set of pupil phase functions having the property to cancel the Strehl ratio (SR) of an optical system were identified and discussed. The similarity in reasoning and the necessary conditions for these analytical functions to fit the requirements of a PMC are exposed in section 2. Based on numerical simulations, a few examples are depicted in section 3, illustrating the achievable performance and, when no trivial solutions exist, giving hints for more robust optimization procedures. A brief conclusion is provided in section 4.

# 2 PRINCIPLE

Let us consider the case of a coronagraphic telescope of focal length $F$ and diameter $D = 2R$, equipped with a phase plate in its image plane and a classical Lyot stop. All employed coordinate systems and scientific notations are defined in Figure 1. For the sake of simplicity, we assume a magnification factor of one between the input pupil and Lyot stop planes, and identical focal lengths for all focusing and collimating optics.

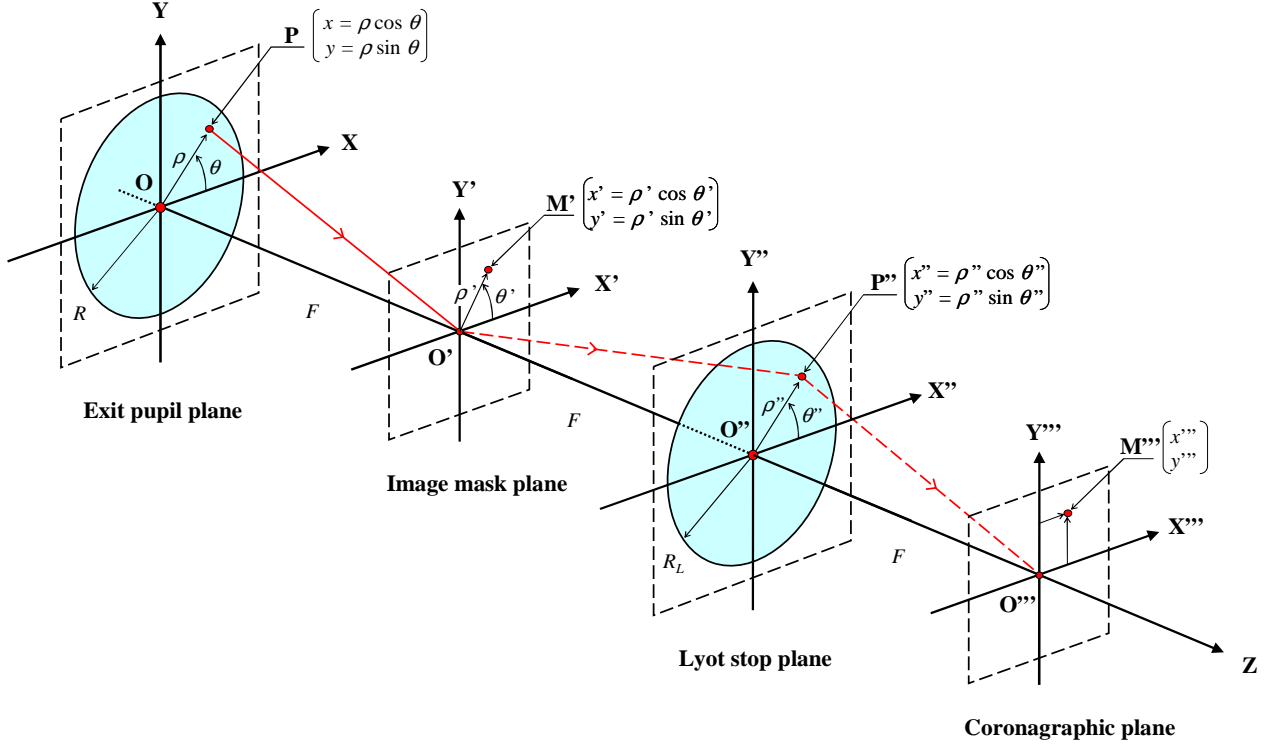

Figure 1: Coordinate systems and scientific notations.

## 2.1 Strehl ratio of an optical system

Following Ref. [11], § 2, the Strehl ratio (SR) achieved by an optical system writes in Cartesian coordinates $x, y$:

$$SR = \left| \iint_{\substack{Pupil \\ aperture}} t(x,y) \exp[2i\pi\Delta(x,y)/\lambda] \, dxdy \, \bigg/ \iint_{\substack{Pupil \\ aperture}} t(x,y) \, dxdy \right|^2, \quad (1)$$

where $t(x,y)$ and $\Delta(x,y)$ respectively are the transmission maps in the input pupil plane OXY, both in amplitude and phase, and $\lambda$ the wavelength of light assumed to be monochromatic. Eq. 1 can also be expressed as function of polar coordinates $\rho$ and $\theta$ in the input pupil plane (see Figure 1):

$$SR = |A'(0,0)|^2 / \pi R^2, \text{ where:} \qquad A'(0,0) = \iint_{\substack{Pupil \\ aperture}} t(\rho,\theta) \exp[2i\pi\Delta(\rho,\theta)/\lambda] \rho \, d\rho d\theta, \quad (2)$$

where $A'(0,0)$ stands for the complex amplitude radiated at the center of the telescope image plane O'X'Y', and $R$ the radius of its input pupil. Assuming as in Ref. [11] that functions $t(\rho,\theta)$ and $\Delta(\rho,\theta)$ are either centro-symmetric (i.e. radially invariant) or azimuthally invariant, one may write:

$$t(\rho,\theta) = B_D(\rho)\, t_r(\rho)\, t_a(\theta) \quad \text{and} \quad \Delta(\rho,\theta) = \delta_r(\rho) + \delta_a(\theta), \tag{3}$$

with $B_D(\rho)$ the "pillbox" function of diameter $D$ representing the telescope input pupil, and $t_r$, $t_a$, $\delta_r$, $\delta_a$ are apodizing functions of a single variable only. Hence from Eqs 2-3 a necessary condition for the system to generate central extinction (i.e. $A'(0,0) = 0$) shall be:

$$A'_r(0,0) = \int_0^{+\infty} B_D(\rho)\, t_r(\rho) \exp[2i\pi\delta_r(\rho)/\lambda]\, \rho\, d\rho = 0 \quad \text{or:} \tag{4a}$$

$$A'_a(0,0) = \int_0^{2\pi} t_a(\theta) \exp[2i\pi\delta_a(\theta)/\lambda]\, d\theta = 0. \tag{4b}$$

The main purpose of Ref. [11] was to provide the reader with a set of analytical phase functions $\delta_r(\rho)$ and $\delta_a(\theta)$, respectively satisfying to the conditions $A'_r(0,0) = 0$ and $A'_a(0,0) = 0$ in the case when $t_r(\rho) = 1$ and $t_a(\theta) = 1$. Some of them will be reutilized in the present study.

### 2.2 Defining the Lyot Strehl ratio (LS)

Reasoning in a similar manner we may define an "amplitude Lyot Strehl ratio" $A''(0,0)$, being an estimator of the central irradiance at the Lyot stop of the coronagraph, and postulate that a necessary condition for an image plane phase plate to produce it writes as:

$$A''(0,0) = A''_r(0,0) \times A''_a(0,0) = \int_0^{+\infty} \hat{B}_D(\rho')\, B'_f(\rho') \exp[2i\pi\dot{\delta}'_r(\rho')/\lambda]\, \rho'\, d\rho' \times \int_0^{2\pi} \exp[2i\pi\dot{\delta}'_a(\theta')/\lambda]\, d\theta' = 0, \tag{5}$$

where:
- $\rho'$ and $\theta'$ are polar coordinates in the image plane O'X'Y' (see Figure 1),
- $f$ is the useful diameter of the phase plate located at the image plane and $B'_f(\rho')$ its pillbox function,
- $\dot{\delta}'_r(\rho')$ and $\dot{\delta}'_a(\theta')$ respectively are the radial and azimuthal phase functions added through the phase plate,
- $\hat{B}_D(\rho')$ is the radial profile of the Fourier transform of the input pupil function $B_D(\rho)$ [1]:

$$\hat{B}_D(\rho') = \frac{2 J_1(\pi D \rho'/\lambda F)}{\pi D \rho'/\lambda F}, \tag{6}$$

for an unobstructed circular pupil, and $J_1$ is the type-J Bessel function at the first order. Eq. 5 is however not sufficient to extinguishing the full Lyot stop area, nor to produce a deep null at the centre of the working image plane of the PMC.

### 2.3 Defining the Coronagraphic Strehl ratio (CS)

Still referring to the coordinate systems of Figure 1, the general expression of the complex amplitude distribution in the Lyot plane is expressed in polar coordinates $\rho''$ and $\theta''$:

$$A''(\rho'',\theta'') = \int_0^{+\infty}\int_0^{2\pi} \hat{B}_D(\rho')\, B'_f(\rho') \exp[2i\pi\dot{\delta}'_r(\rho')/\lambda] \exp[2i\pi\dot{\delta}'_a(\theta')/\lambda] \exp[2i\pi\rho'\rho''\cos(\theta''-\theta')/\lambda F]\, d\theta'\, \rho'\, d\rho'. \tag{7}$$

---

[1] Note that $\hat{B}_D(\rho')$ is not the Fourier transform of $B_D(\rho)$. Instead, these functions are the radial profiles of two bi-dimensional, axis symmetric distributions who are Fourier-transformed each other.

We then define the "coronagraphic Strehl ratio" $A'''(0,0)$ as: $\quad A'''(0,0) = \int_0^{+\infty}\int_0^{2\pi} B''_L(\rho'') A''(\rho'',\theta'') d\theta'' \rho'' d\rho''$, (8) where $B''_L(\rho'')$ is the pillbox function defining the area of the Lyot stop, assumed to be circular and of diameter $D_L = 2R_L$ (see Figure 1). Simply inserting relation 7 into Eq. 8 and permuting the double integrals over coordinates ($\rho'$, $\theta'$) and ($\rho''$, $\theta''$) allows determining a reduced expression for $A'''(0,0)$:

$$A'''(0,0) = A'''_r(0,0) \times A''_a(0,0) = \int_0^{+\infty} \hat{B}_D(\rho') \hat{B}''_L(\rho') B'_f(\rho') \exp\left[2i\pi \dot{\delta}'_r(\rho')/\lambda\right] \rho' d\rho' \times \int_0^{2\pi} \exp\left[2i\pi \dot{\delta}'_a(\theta')/\lambda\right] d\theta', \quad (9)$$

with $\hat{B}''_L(\rho')$ the Fourier transform of the Lyot stop estimated in the same way as indicated in relation 6.

### 2.4 Summary

Using the same mathematical approach as in Ref. [11] and applying it to a coronagraphic telescope, we previously introduced the notions of "Lyot" and "coronagraphic" Strehl ratios and used them for defining necessary conditions for achieving null irradiance at the centre of both the Lyot and image planes of a phase plate coronagraph. It was demonstrated that:

1) In the case of a purely azimuthal phase function $\dot{\delta}'_a(\theta')$, the sole condition for ensuring central extinction at the Lyot and PMC image planes writes as:

$$A''_a(0,0) = \int_0^{2\pi} \exp\left[2i\pi \dot{\delta}'_a(\theta')/\lambda\right] d\theta' = 0. \quad (10)$$

2) The case of a radial phase function $\dot{\delta}'_r(\rho')$ is a bit less easy, since the both following conditions have to be fulfilled simultaneously:

$$A''_r(0,0) = \int_0^{+\infty} \hat{B}_D(\rho') B'_f(\rho') \exp\left[2i\pi \dot{\delta}'_r(\rho')/\lambda\right] \rho' d\rho' = 0, \quad \text{and:} \quad (11a)$$

$$A'''_r(0,0) = \int_0^{+\infty} \hat{B}_D(\rho') \hat{B}''_L(\rho') B'_f(\rho') \exp\left[2i\pi \dot{\delta}'_r(\rho')/\lambda\right] \rho' d\rho' = 0, \quad (11b)$$

with $\hat{B}_D(\rho')$ and $\hat{B}''_L(\rho')$ being defined as in Eq. 6.

3) Finally, most of the mathematical properties of functions $\dot{\delta}'_a(\theta')$ and $\dot{\delta}'_r(\rho')$ highlighted in Ref. [11] remain applicable in the image plane of the coronagraph, especially for what concerns additions, rescaling in truncated apertures, and spatial combination of phase functions.

The remainder of this study will now be essentially focused at providing a few illustrative examples to the here above properties.

## 3  NUMERICAL SIMULATIONS AND ACHIEVABLE PERFORMANCE

In this section will be reviewed different types of phase functions satisfying to the necessary conditions 10-11, and their performance will be discussed. They are the optical vortex (§ 3.1), azimuthal cosine-modulated (§ 3.2), and circular grating phase functions (§ 3.3).

## 3.1 Vortex phase mask (VPM)

From its initial proposal in the field of coronagraphy [7-8] to on-sky validation [9] the optical vortex has been recognized as one of the most efficient types of PMC. Their phase function is classically defined as:

$$\varphi'_a(\theta') = 2\pi \dot{\delta}'_a(\theta')/\lambda = k\theta', \qquad (12)$$

where $k$ is an integer defined as the topological charge of the optical vortex. It can be verified that this function is a trivial solution of the necessary condition expressed in Eq. 10, therefore it should generate central extinctions at both Lyot and image planes of the PMC. The vortex function is illustrated in Figure 2, showing a three-dimensional view of a phase plate generating an optical vortex. The figure is a good illustration of the technical difficulties that may arise for fabricating the VPM, at least by means of a phase plate or of a deformable mirror:
- At the origin (x'=0, y'=0) the phase mask is undefined mathematically (it could take any value comprised between 0 and $2\pi$). This ambiguity can be raised by drilling a central hole at the phase plate,
- However a vortex phase function must necessarily present a $2\pi$ phase jump across the X' (or real) axis. The resulting discontinuity strongly limits the choice of manufacturing technologies. So far the best results were obtained by use of sub-wavelength gratings [9], not with deformable mirrors or phase plates.

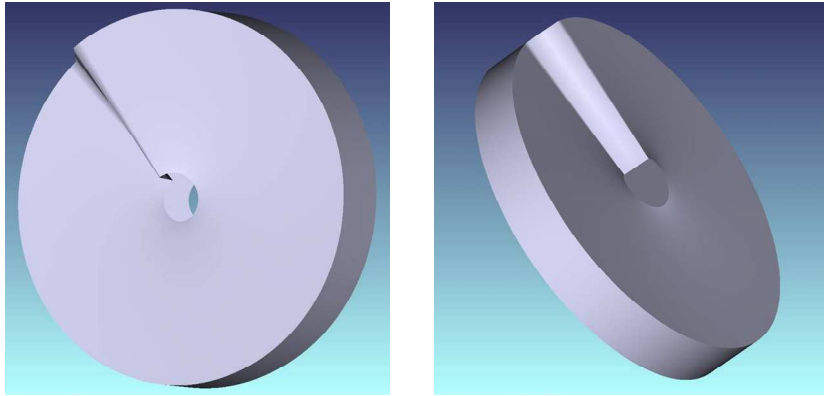

Figure 2: CAD views of a phase plate generating an optical vortex.

The achievable performance of a vortex PMC is illustrated on the left side of Figure 3: The upper left panel shows the irradiance distribution produced at the Lyot stop plane X"Y". One can see that most of the starlight is concentrated at a ring corresponding to the input pupil of the telescope. Very low light levels are observed inside the ring, and the presence of the Lyot stop (here in yellow dashed lines) slightly undersized with respect to the telescope pupil usually reduces the starlight to a negligible level in the coronagraphic plane, as shown in the lower right vignette. Figure 3 also illustrates the evolution of the irradiance distributions in the Lyot and coronagraphic planes as function of the angular Field of view (FoV), expressed in terms of lateral displacement x' in the focal plane of the telescope. Numerical simulations were carried out using the Matrix Fourier transform (MFT) algorithm described in Ref. [12] with pupil and image plane sampling of 4096 x 4096. Similar results were also obtained with the Fast Fourier transform (FFT). The other numerical parameters were set to $D = 10$ m, $F = 200$ m and $\lambda = 0.5$ µm. The diameter of the Lyot stop has been set to $D_L = 0.8D$.

It is well known that the presence of central telescope obstruction severely hampers the performance of a vortex coronagraph [13-14]. This effect is illustrated on the right side of Figure 3 for a central obscuration ratio equal to $\tau = 0.2$. In the upper left panel, one can distinctively see that:
- Faint diffraction rings are appearing near the central obscuration area in the Lyot stop plane X"Y". They are located outside the obstructing circle and thus difficult to block with the help of an additional Lyot mask. Therefore a significant fraction of unwanted starlight shall be diffracted at the coronagraphic image plane.

- This is confirmed by the coronagraph response shown in the lower right vignette. However the general aspect of this nulling response function remains identical to an optical vortex, with its characteristic central null surrounded by a bright ring of light. This is in full agreement with the theoretical analysis of section 2, from which it can be inferred that central nulls generated by azimuthal phase functions are not affected by axis-symmetric obscurations (Eqs. 5 and 9).

It can be concluded that the necessary condition (10) for achieving central nulls at both the Lyot and image planes of the PMC is not a sufficient condition for total extinction of the central star observed by the coronagraph – as could be expected. Nevertheless, it remains of high interest to study alterative types of phase mask functions suitable to implementation into a PMC.

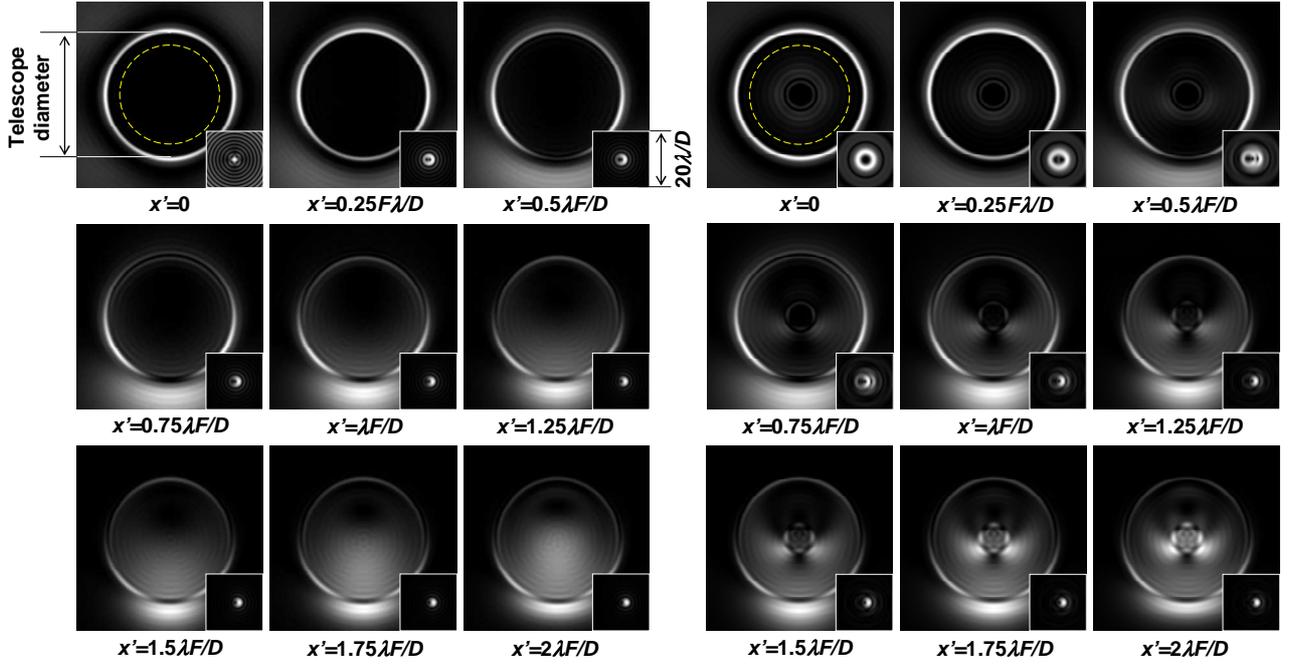

Figure 3: Irradiance maps generated by a VPM of topological charge 4 in the Lyot stop plane X"Y" of the coronagraph, for different positions x' in the FoV of the instrument. The bottom left vignettes show the corresponding nulling responses in the coronagraphic plane X'"Y'". The actual diameter of the Lyot stop is indicated in yellow dashed lines. Left side: case of unobstructed telescope pupil. Right side: case of central obscuration ratio $\tau = 0.2$.

### 3.2 Azimuthal cosine-modulated (ACM) phase functions

#### 3.2.1 Definition

The azimuthal cosine-modulated (ACM) phase function is another possible solution of the necessary condition (10). Mathematically, it is defined as [11]:

$$\varphi'_a(\theta') = 2\pi \dot{\delta}'_a(\theta')/\lambda = z_n \cos(k\theta'), \quad (13)$$

where $k$ is the angular frequency of the function and $z_n$ is the n[th] zero of the type-J Bessel function $J_0(z)$. An example of phase plate generating such functions is shown in Figure 4 (with $k = 15$ for the sake of illustration). One of the major advantages of this ACM phase plate clearly appears here: except at the origin O'; no discontinuity nor slope break exist at the surface of the phase plate. The indetermination around point O' can itself be eliminated by means of a central hole as shown in the figure.

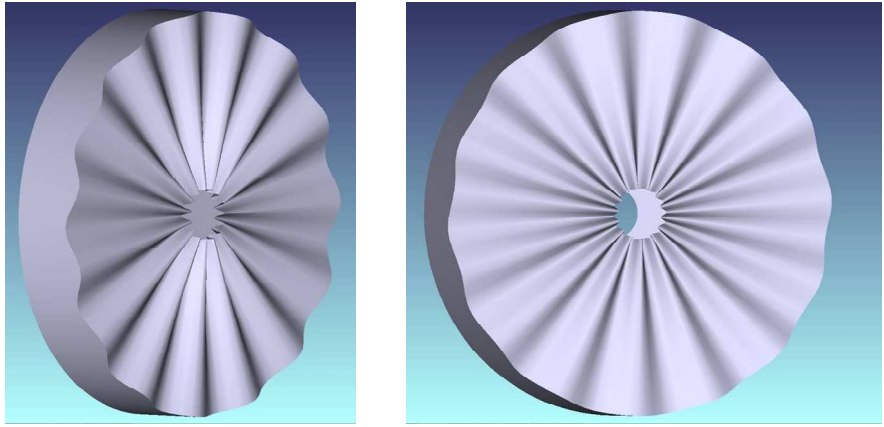

Figure 4: CAD views of a phase plate generating an ACM phase function of angular frequency 15.

### 3.2.2 *Comparison with optical vortex*

A qualitative comparison between the ACM and optical vortex phase functions is presented in Figure 5. In the left side is firstly illustrated the case of an unobstructed telescope pupil. The angular frequency of the ACM function is set equal to 4. The irradiance map formed at the Lyot stop plane is shown on the upper left panel. As in Figure 3 most of starlight is diffracted outside the pupil. However the external and uniform ring of light is now replaced by eight symmetric diffraction lobes. More generally, other numerical simulations – not shown here – demonstrate that an ACM function of $k$ angular frequency will generate $2k$ external diffraction lobes. Figure 5 also displays irradiance distribution maps in the Lyot and coronagraphic planes as function of the lateral displacement x' in the telescope FoV (left side), and the effect of a telescope central obscuration ratio equal to $\tau = 0.2$ (right side). These numerical simulations show that the ACM and vortex phase functions globally share the same behavior.

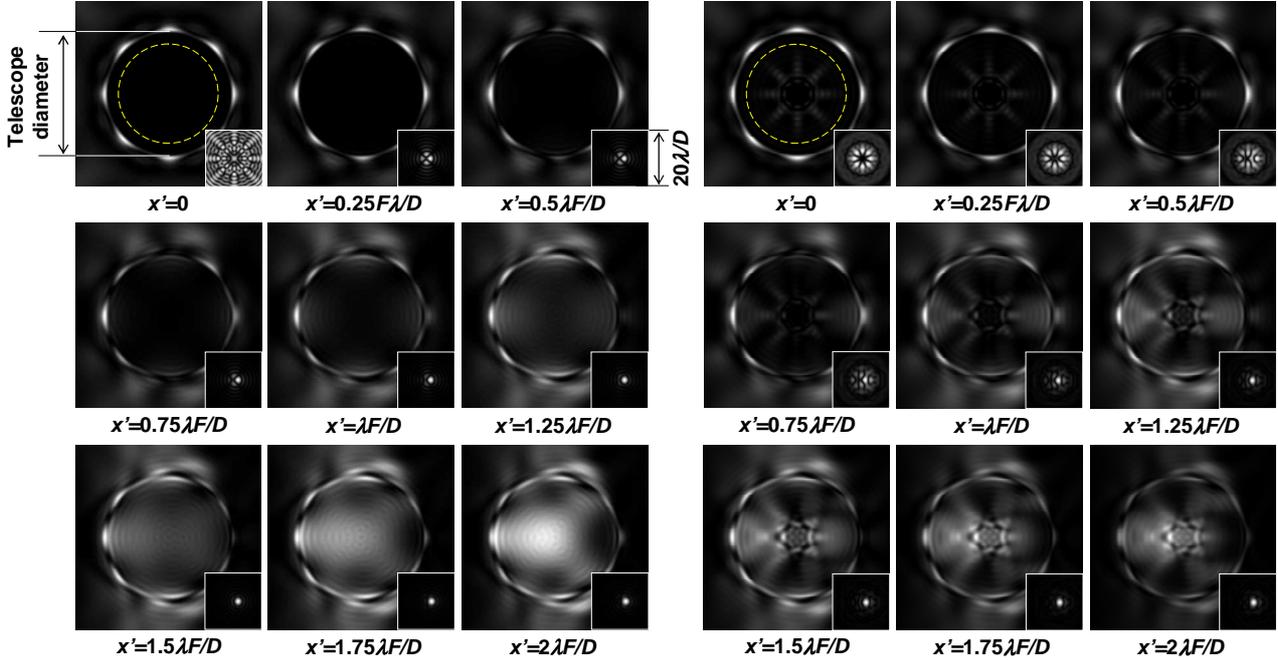

Figure 5: Same illustration as in Figure 3 for the case of an ACM phase function of angular frequency 4.

A more quantitative performance analysis is illustrated in Figure 6, showing different views of the Inner working angle (IWA) curves achievable with both types of phase plates. It must be noted that both the throughput performance and sensitivity to pointing errors can be deduced from those curves. The considered cases and obtained results are summarized in Table 1. In particular, they reveal that:

1) The achieved null rates by the vortex and ACM functions are rather good and all inferior to $5 \cdot 10^{-7}$. They are actually estimated as the ratio of luminous energy reaching the coronagraph image plane and integrated over a circle of radius $20\lambda/D$, with respect to total energy collected by the telescope. Vortex performance looks slightly better without exhibiting decisive advantage. Here and for purpose of comparison the differences between the null values are more significant than their absolutes values, the latter being currently limited by the allowed pupil and image grids sampling, and resulting computing times.

2) Estimated IWAs also show the same tendency. For example the vortex function of charge 2 attains an IWA of $0.927\lambda/D$, to be compared with $0.984\lambda/D$ for an ACM function of angular frequency 2. One may remark however that the ACM curves are increasing more rapidly above $2\lambda/D$, thus providing small flux improvement for extra-solar planets very close to the central star.

3) Not surprisingly, the performance of the coronagraph collapses when the collecting telescope shows a central obscuration. Both types of functions are affected in a similar manner. Methods to recover coronagraphic performance in presence of pupil obscurations are discussed in Refs. [13-14].

4) Finally, it turns out that a slight central obscuration of the Image plane mask (here of diameter $2\lambda/D$, or 10 µm with the employed numerical parameters) does not significantly affect the nulling rate and IWA performance. This is perhaps the main advantage of ACM phase functions because their central discontinuity is eliminated and they can be generated by a phase plates or by a deformable mirror much more easily[1].

It can then be concluded that ACM phase functions, despite of slightly worst performance, represent a promising alternative to the now classical optical vortex into a PMC. In the next sub-section will now be examined the second family of phase functions defined in § 2 that are all axis-symmetric, solely depending on radial coordinate $\rho'$.

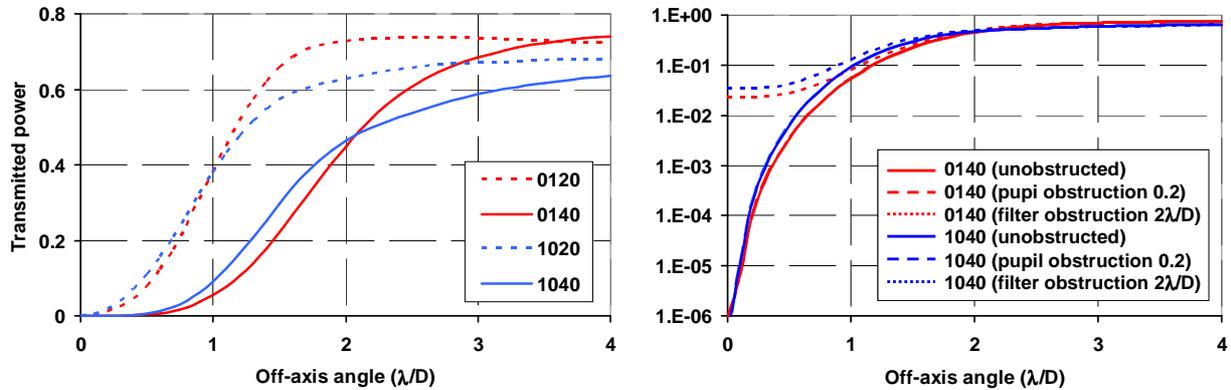

Figure 6: IWA curves calculated for different types of nulling phase functions. Left side: Comparison between vortex functions 1020 and 1040 and ACM functions 0120 and 0140. Right side: Effect of a central telescope obscuration ratio $\tau = 0.2$ and of a central filter obscuration equal to $2\lambda/D$ (or 10 µm with the employed numerical parameters). These two last curves are visually undistinguishable from the unobstructed case.

---

[1] Note that other technologies could be employed for generating phase discontinuities, see e.g. Ref. [15].

Table 1: Summary of numerical simulations for azimuthal phase functions.

| Case | Phase functions | Code (*) | Obstructions | Central null | IWA ($\lambda/D$) | Illustration |
|---|---|---|---|---|---|---|
| 1 | Vortex of topological chaege 2 | 1020 | None | 1.5E-07 | 0.927 | Fig. 6a |
| 2 | Vortex of topological chaege 4 | 1040 | None | 2.1E-07 | 1.550 | Figs. 3a, 6a, 6b |
| 3 | ACM of angular frequency 2 | 0120 | None | 2.0E-07 | 0.984 | Fig. 6a |
| 4 | ACM of angular frequency 4 | 0140 | None | 4.2E-07 | 1.822 | Figs. 5a, 6a, 6b |
| 5 | Vortex of topological chaege 4 | 1040 | Input pupil, $\tau = 0.2$ | 2.2E-02 | 1.452 | Figs. 3b, 6b |
| 6 | ACM of angular frequency 4 | 0140 | Input pupil, $\tau = 0.2$ | 1.5E-02 | 1.731 | Figs. 5b, 6b |
| 7 | Vortex of topological chaege 4 | 1040 | Phase mask, $2\lambda/D$ | 3.1E-07 | 1.552 | Fig. 6b |
| 8 | ACM of angular frequency 4 | 0140 | Phase mask, $2\lambda/D$ | 6.1E-07 | 1.822 | Fig. 6b |

(*) From Ref. [11]

### 3.3 Circular phase gratings (CPG)

#### 3.3.1 Optimization methodology

In Ref. [11], the CPG was identified as a solution of Eq. 4a, thus generating a central null at the image plane in the case when $t_r(\rho) = 1$. The typical shape of this phase function is illustrated in Figure 7 and its mathematical expression writes as:

$$\varphi_r(\rho) = 2\pi \delta_r(\rho)/\lambda = z_n \cos(2\pi k \rho), \tag{14}$$

where $k$ is the integer number of radial periods on the grating aperture and $z_n$ is the same as in Eq. 13. Therefore it is of some interest to verify if similar phase functions of the form $\varphi'_r(\rho) = a' \cos(2\pi f' \rho)$, where $a'$ and $f'$ are real numbers, should serve as the phase mask of a PMC. If such solutions exist, they should necessarily satisfy to the equations 11 derived in sub-section 2.4. The latter may be rewritten as:

$$A''_r(0,0) = \frac{2\lambda F}{\pi D} \int_0^{R_F} J_1(\pi D \rho'/\lambda F) \exp[i\varphi'_r(\rho')] d\rho' = 0, \quad \text{and:} \tag{15a}$$

$$A'''_r(0,0) = \frac{4\lambda^2 F^2}{\pi^2 DL} \int_0^{R_F} \frac{J_1(\pi D \rho'/\lambda F) J_1(\pi L \rho'/\lambda F)}{\rho'} \exp[i\varphi'_r(\rho')] d\rho' = 0, \tag{15b}$$

where $R_F$ is the spatial radius of the phase mask assumed to circular and of limited size, and the explicit forms of $\hat{B}_D(\rho')$ and $\hat{B}''_L(\rho')$ were introduced from Eq. 6. However those equations (even considered separately) do not admit simple analytical solutions a priori. Herein we shall only look for these solutions numerically, exploring the ($a'$, $f'$) space and determining the minima of the two following criteria:

$$C_1 = |A''_r(0,0)|^2 \qquad \text{for null irradiance at the Lyot plane centre only, and:} \tag{16a}$$

$$C_2 = \sqrt{\frac{4\lambda^2 F^2}{\pi^2 DL}|A''_r(0,0)|^2 + |A'''_r(0,0)|^2} \qquad \text{for null irradiances at both Lyot and coronagraphic planes centres.} \tag{16b}$$

The results of this elementary optimization process are summarized in Table 2 and discussed in the following sub-section. The explored ranges for parameters ($a'$, $f'$) were limited to $0 \leq a' \leq 10$ and $0 \leq f' \leq 10$.

#### 3.3.2 Examples of results

The most significant numerical results are illustrated in Figure 8. We basically considered three cases:

- <u>Case 1</u>: Only central nulls at the Lyot stop plane are search for, thus minimizing criterion $C_1$. The obtained numerical values for $a'$, $f'$, and the achieved nulling rate and IWA are given in Table 2. One can see that irradiance actually cancels at the Lyot plane centre, but not over the whole telescope pupil area. Practically the maximal radius of the darkened zone can be estimated as $D_L = 0.3\ D$, over which bright diffraction rings generated by the CPG are visible. Fitting the Lyot stop to this limit (the red circle in Figure 8), does not provides an efficient nulling rate (Table 2 and blue curve on the lower panel). This result confirms that $C_1$ criterion alone is a necessary but not sufficient condition for achieving full extinction of the central star, as already pointed out in § 3.1.

- <u>Case 2</u>: Here the $C_2$ criterion was minimized, leading to a best compromise between the extinction rates at the centres of both Lyot stop and final image planes. The Lyot stop diameter is limited at $D_L = 0.7\ D$ in order to remove an external ring of stray radiation. The results are shown in the central column of the figure, distinctively exhibiting central nulls at both planes simultaneously. It must be noticed however that the achieved starlight extinction is particularly disappointing (6.1 $10^{-3}$ integrated over a $20\lambda/D$ circular area in the final coronagraphic plane). This is essentially due to the inner diffraction ring in the Lyot plane, clearly visible at the near vicinity of its central null. Hence it can be concluded that, like $C_1$, the $C_2$ criterion is not sufficient for realizing total extinction of the central star.

- <u>Case 3</u>: Starting from case 2, we simply eliminated the inner diffraction ring by adding a central obscuration equal to $0.3\ D$ in the Lyot stop plane. As expected the central is preserved in the coronagraphic plane and the nulling rate is improved, but only by a modest factor (around 6).

At the end of this preliminary study, one may conclude that, contrarily to the azimuthal cosine phase functions, circular phase gratings are not very promising for insertion into a PMC. However this conclusion may be moderated somewhat, because:

- Only a limited range for parameters ($a'$, $f'$) was explored.
- The influence of other parameters was not studied: It would be interesting in particular to employ a more efficient optimization algorithm, and adding as variables the diameters of the phase filter and of the Lyot stop, and possibly a phase-shift inside the cosine function.
- Discontinuous or frequency-varying CPG phase functions could also be optimized with similar procedures.

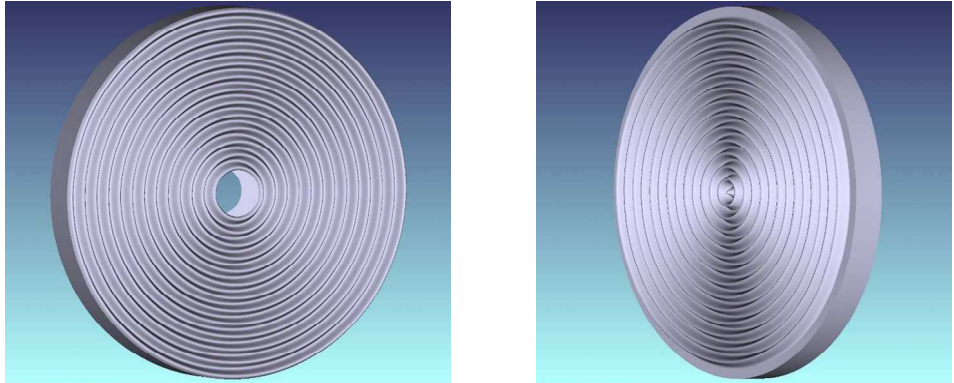

Figure 7: CAD views of a CPG radial phase function with $f' = 15$ (the central hole on the left panel is for illustration purpose only).

Table 2: Summary of numerical simulations for radial phase functions.

| Case | Phase functions | Amplitude $a'$ | Frequency $f'$ | Optimization criterion | Lyot stop geometry | Central null | IWA ($\lambda/D$) |
|---|---|---|---|---|---|---|---|
| 1 | Circular grating | 8.192 | 7.595 | $C_1$ | $D_L = 0.3\ D$ | 1.4E-05 | 0.782 |
| 2 | Circular grating | 2.445 | 9.706 | $C_2$ | $D_L = 0.7\ D$ | 6.1E-03 | 0.316 |
| 3 | Circular grating | 2.445 | 9.706 | $C_2$ | $D_L = 0.7\ D$, $t_L = 0.3$ | 1.1E-03 | 0.395 |

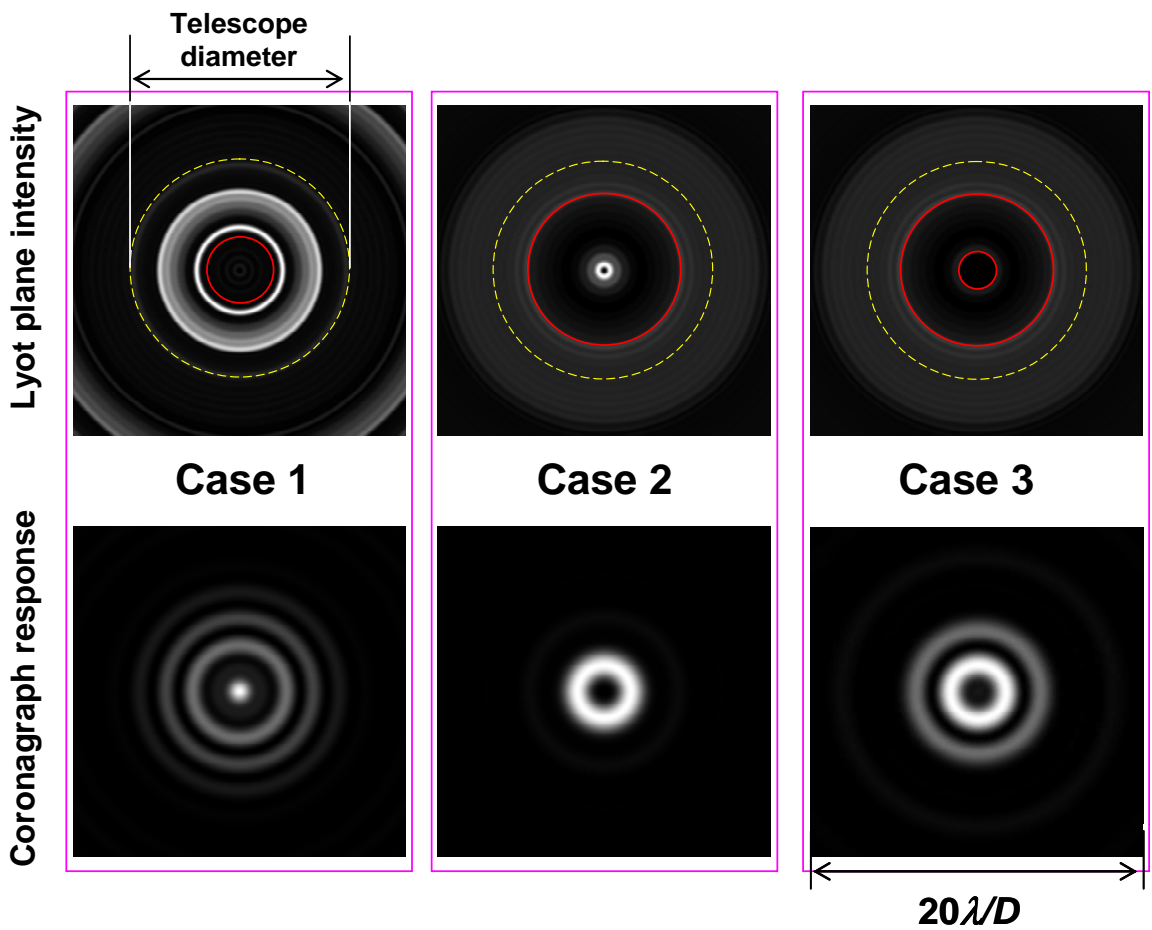

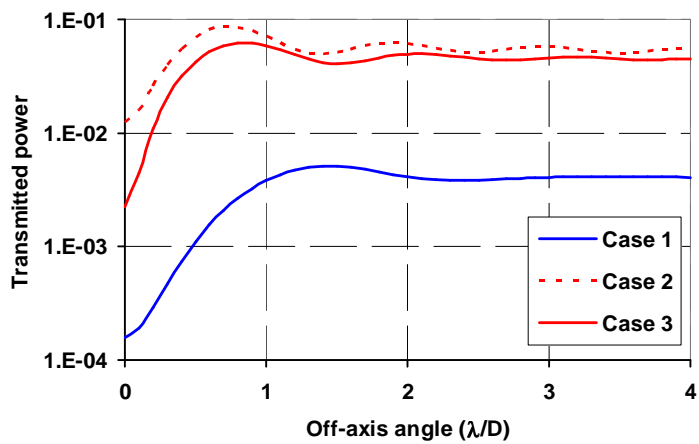

Figure 8: Irradiance maps generated by CPG phase functions in the Lyot stop plane of the coronagraph (upper panels) and the corresponding nulling responses in the coronagraphic plane (central panels). The telescope diameter is indicated in yellow dashed lines, while solid red lines show the actual obscurations of the Lyot stop. IWA curves are displayed on the lower panel.

## 4   CONCLUSION

The aim of this paper was to identify new types of phase functions suitable to a phase-mask coronagraph, where phase plates located at an intermediate image plane are used to spread the starlight outside the exit pupil diameter of the telescope, thus enabling the detection of extra solar planets. For that purpose, we started from a set of continuous analytical functions described in Ref. [11], having the property to cancel the Strehl ratio of an uniformly illuminated optical system. The necessary conditions for these analytical functions to fit the requirements of a PMC were defined in section 2. Based on numerical simulations, a few examples of azimuthally-invariant or centro-symmetric phase functions were discussed in section 3, including azimuthal cosine-modulated phase functions and circular phase gratings. It was also shown that the herein necessary conditions are sometimes insufficient for achieving complete extinction of the central star in the final image plane, and should require using more robust optimization procedures.

From a more practical point of view, it seems that the azimuthal cosine-modulated phase functions represent a good alternative to the well-known "vortex" phase ramps, because they exhibit comparable performance and can be manufactured more easily as phase plates without discontinuities.

FH acknowledges funding help from the French "Action spécifique haute résolution angulaire" (ASHRA) managed by CNRS-INSU.